\documentclass[%
reprint,
superscriptaddress,
floatfix,
]{revtex4-1}

\usepackage{amssymb,amsmath}
\usepackage{graphicx}
\usepackage[tight,footnotesize]{subfigure}
\usepackage{dcolumn}
\usepackage{bm}
\usepackage[utf8]{inputenc}
\usepackage{gensymb}
\usepackage{units}

\def\mathclap#1{\text{\hbox to 0pt{\hss$\mathsurround=0pt#1$\hss}}} 
\newcommand{\msub}[1]{_{\text{#1}}}
\newcommand{\musb}[1]{_{\text{#1}}}

\newcommand{\refF}[1]{Fig.\,\ref{#1}}
\newcommand{\refT}[1]{Tab.\,\ref{#1}}
\newcommand{\refE}[1]{Eq.\,\ref{#1}}

\DeclareFontFamily{U}{euc}{}
\DeclareFontShape{U}{euc}{m}{n}{<-6>eurm5<6-8>eurm7<8->eurm10}{}%
\DeclareSymbolFont{AMSc}{U}{euc}{m}{n} 
\DeclareMathSymbol{\umu}{\mathord}{AMSc}{"16}

\begin{document}
\preprint{APS/123-QED}

\title{Superheating in coated niobium}

\author{T. Junginger} 
\email{Tobias.Junginger@helmholtz-berlin.de}
\affiliation{Helmholtz-Zentrum Berlin fuer Materialien und Energie (HZB), Germany}
\affiliation{TRIUMF Canada's National Laboratory for Particle and Nuclear Physics, Vancouver}
\author{W. Wasserman}
\affiliation{U. of British Columbia, Vancouver, Canada}
\author{R. E. Laxdal}
\affiliation{TRIUMF Canada's National Laboratory for Particle and Nuclear Physics, Vancouver}

\date{\today}

\begin{abstract}
Using muon spin rotation it is shown that the field of first flux penetration $H\msub{entry}$ in Nb is enhanced by about \unit[30]{\%} if coated with an overlayer of Nb$_3$Sn or MgB$_2$. This is consistent with an increase from the lower critical magnetic field $H\musb{c1}$ up to the superheating field $H\msub{sh}$ of the Nb substrate. In the experiments presented here coatings of Nb$_3$Sn and MgB$_2$ with a thickness between 50 and \unit[2000]{nm} have been tested. $H\msub{entry}$ does not depend on material or thickness. This suggests that the energy barrier at the boundary between the two materials prevents flux entry up to $H\msub{sh}$ of the substrate. A mechanism consistent with these findings is that the proximity effect recovers the stability of the energy barrier for flux penetration, which is suppressed by defects for uncoated samples. Additionally, a low temperature baked Nb sample has been tested. Here a \unit[6]{\%} increase of $H\msub{entry}$ was found, also pushing $H\msub{entry}$ beyond $H\msub{c1}$.     

\end{abstract}


\maketitle 

\section{Introduction}
The use of multilayers of superconductors with a larger critical temperature $T\msub{c}$ on top of Nb to increase the accelerating gradient $E\msub{acc}$ of superconducting radiofrequency (SRF) cavites has first been proposed by Gurevich \cite{Gurevich_mulitlayers}. He suggested a structure with interlaying insulating layers to prevent flux penetration in the Nb substrate. In \cite{kubo2014radio} this structure has been studied within London theory and it was concluded that also a layered structure without insulators can potentially yield a larger field of first flux penetration $H\msub{entry}$ compared to an uncoated substrate. Highest $E\msub{acc}$ with Nb cavities is achieved by low temperature baking at \unit[120]{$\degree$C} in vacuum for \unit[48]{h} following wet acid chemistry. Low energy muon spin rotation measurements have shown that there is a change in the Meissner screening of low temperature baked Nb, i.e. a depth dependent mean free path \cite{Romanenko_muSR}. The material can therefore be considered as an effective multilayer system. 

Kubo proposed two mechanisms that link the change in Meissner screening to the enhanced $H\msub{entry}$ \cite{kubo2016multilayer}. (1) A counter current flow at the boundary between the two superconductors suppresses the surface current and therefore enhances the theoretical field limit. This yields a maximum $H\msub{entry}$ above the individual superheating field of the substrate and the coating. (2) There is a second energy barrier at the boundary between the two superconductors. It has to be noted that (1) requires that both materials can be operated in a superheated state. This is rather unlikely, due to vortex penetration at defects. In fact, experimental results suggest that technical superconductors can generally not be superheated. An exception is \unit[120]{$\degree$C} baked Nb as used for SRF cavities. However also for this case the maximum field values are below the prediction of the effective two layer model. Kubo suggested that this system should be described by an infinite number of thin superconductors continuously piled up on a substrate. Therefore, the theoretical field limit for this structure, especially with defects and surface roughness, is hard to estimate \cite{kubo2015field}. 

Checchin in \cite{checchin2016physics} introduced a sigmoidal function for the Ginzburg Landau parameter $\kappa$, representing the depth dependent mean free path. He solves the normalized one-dimensional Ginzburg-Landau equations to estimate the forces acting on a vortex, and then looks at the Bean-Livingston barrier \cite{Bean_Livingston} created by the forces from the Meissner current and an image vortex introduced to fulfill the boundary condition at the superconductor-vacuum (SV) interface. 
He finds that the energy barrier is enhanced for a sigmoidal compared to a constant $\kappa$. This enhancement depends on the distance over which $\kappa$ changes and the thickness of the outer layer. If the latter becomes large compared to the former, the model reduces to an effective bi-layer system with two distinct energy boundaries. \\

\section{Experiment}
All theoretical considerations for $H\msub{entry}$ reviewed above are not restricted to the RF case. In fact, there are several non-intrinsic field limitations of SRF cavities, i.e. field emission, multipacting and premature quench, which in general prevent reaching the intrinsic $H\msub{entry}$. It is therefore beneficial to use a DC method to measure $H\msub{entry}$. For this purpose we have established a muon spin rotation ($\mu$SR) experiment \cite{Junginger_muSR_Overview}. Spin polarized muons with an average stopping distance of \unit[130]{$\mu$m} are implanted one at a time into the sample. When the muon decays (half life=\unit[2.197]{$\mu$s}) it emits a fast decay positron, preferentially along the direction of its spin. 
%
%
By detecting the location of emitted positrons as a function of time with two detectors the spin precession of the muons and therefore magnetic field properties can be inferred through an asymmetry signal
\begin{equation}
	Asy(t)=\frac{N\msub{l}(t)-\alpha N\msub{r}(t)}{N\msub{l}(t)+\alpha N\msub{r}(t)},
	\label{eq:Alpha}
\end{equation}
where $N\msub{l}(t)$ and $N\msub{r}(t)$ are the number of counts in the left and right detector. The parameter $\alpha$ is added to account for detector efficiencies and to remove any bias between the up and down detectors caused by uneven solid angles. In the case where the detector efficiencies are identical, $\alpha$ assumes a value of 1. Samples are placed in a cryostat surrounded by by field-inducing coils. For field penetration measurements, samples are cooled in zero field to below $T\msub{c}$ in a horizontal gas flow cryostat, which allows to reach a base temperature of about \unit[2.5]{K} and then a static magnetic field is applied perpendicular to the initial spin polarization to probe if field has penetrated the sample. Specifically, the asymmetry signal gives information on the volume fraction of the host material sampled by the muon that does not contain magnetic field. This signal can be used to characterize the superconducting state, particularly the transition from Meissner to mixed state. The total asymmetry function is a sum of two terms. The first one is the dynamic Kubo-Tuyabe function \cite{Hayano} with initial asymmetry $a_0$. The second term is a damped oscillating function caused by the penetrated external field:
\begin{eqnarray}
Asy(t)&=&a_0\cdot P_{\rm ZF}^{\rm dyn.}(t) + \\ \nonumber
& & a_1\cdot\exp{\left( -\frac{1}{2}\Delta^2t^2\right) }\cdot \cos{\left( \omega t + \frac{\pi\phi}{180}\right) }
\label{eq:Asy}
\end{eqnarray}
with 
\begin{equation}
\omega=2 \pi \gamma_\mu H\msub{int},
\end{equation}
where $\gamma_\mu$=13.55KHz/G is the gyromagnetic ratio of the muon and $\phi$ the phase. 
The value of $a_0$ compared to its initial low field value is a measure of the volume fraction being in the field free Meissner state. \\


\section{Sample preparation}
All samples were made from fine grain niobium from Tokyo Denkai with a residual resistance ratio (RRR) above 300. In \cite{Junginger_muSR_Overview} a detailed study of geometry and pinning has been carried out. Depending on sample and field geometry, $H\msub{entry}$ measurements can potentially give an artificially larger value due to flux pinning. Careful precautions have been taken to avoid this for the measurements presented here. Ellipsoidal samples with the magnetic field applied along the major axis with the muons implanted at the equator are rather insensitive to pinning and ideally suited for $H\msub{entry}$ measurements. This configuration is used for one Nb$_3$Sn and one Nb sample. For the MgB$_2$ measurements coin samples have been used to simplify the coating procedure. Here the field is applied in the radial direction. In \cite{Junginger_muSR_Overview} it has been shown that this geometry is only slightly more sensitive to pinning than the ellipsoidal geometry. For comparison Nb and Nb$_3$Sn samples of a coin shape have also been produced for this study. 

These coin samples have a thickness of \unit[3]{mm} and are \unit[20]{mm} in diameter. They were cut by water jet from sheets. The prolate ellipsoids were machined to the dimensions of a semi-major axis of \unit[22.9]{mm} and a semi-minor circular cross-section of \unit[6.3]{mm} radius. After machining the samples were treated by buffered chemical polishing (BCP) to remove \unit[100]{$\mu$m} of outer material. Afterwards all samples, except the ones which were used for Nb$_3$Sn coating, have received a \unit[1400]{$\degree$C} annealing at TRIUMF. This treatment is effective in releasing virtually all pinning \cite{Junginger_muSR_Overview}. After the heat treatment the samples received an additional BCP to remove another \unit[30]{$\mu$m} of material. The Nb ellipsoid received also a \unit[120]{$\degree$C} baking in vacuum at TRIUMF after initial testing.

The Nb$_3$Sn coatings were produced by vapor diffusion at Cornell University. This process includes heating up to \unit[1100]{$\degree$C} which also strongly releases pinning. Furthermore, in \cite{Junginger_muSR_Overview} Nb$_3$Sn has been studied in different geometries and it was concluded that pinning is rather weak for this sample. The MgB$_2$ coatings were carried out at Temple University using the Hybrid Physical-Chemical Vapor Deposition (HPCVD) technique. For details about the coating procedures refer to \cite{posen2017nb3sn} and \cite{zeng2002situ}. 

Since all Nb substrates have received a high temperature annealing and pinning is mainly a bulk effect the measurements presented here have no ambiguity concerning $H\msub{entry}$ vs pinning. 

\section{Experimental Results}
\begin{figure}
	\centering
		\includegraphics[width=0.90\columnwidth]{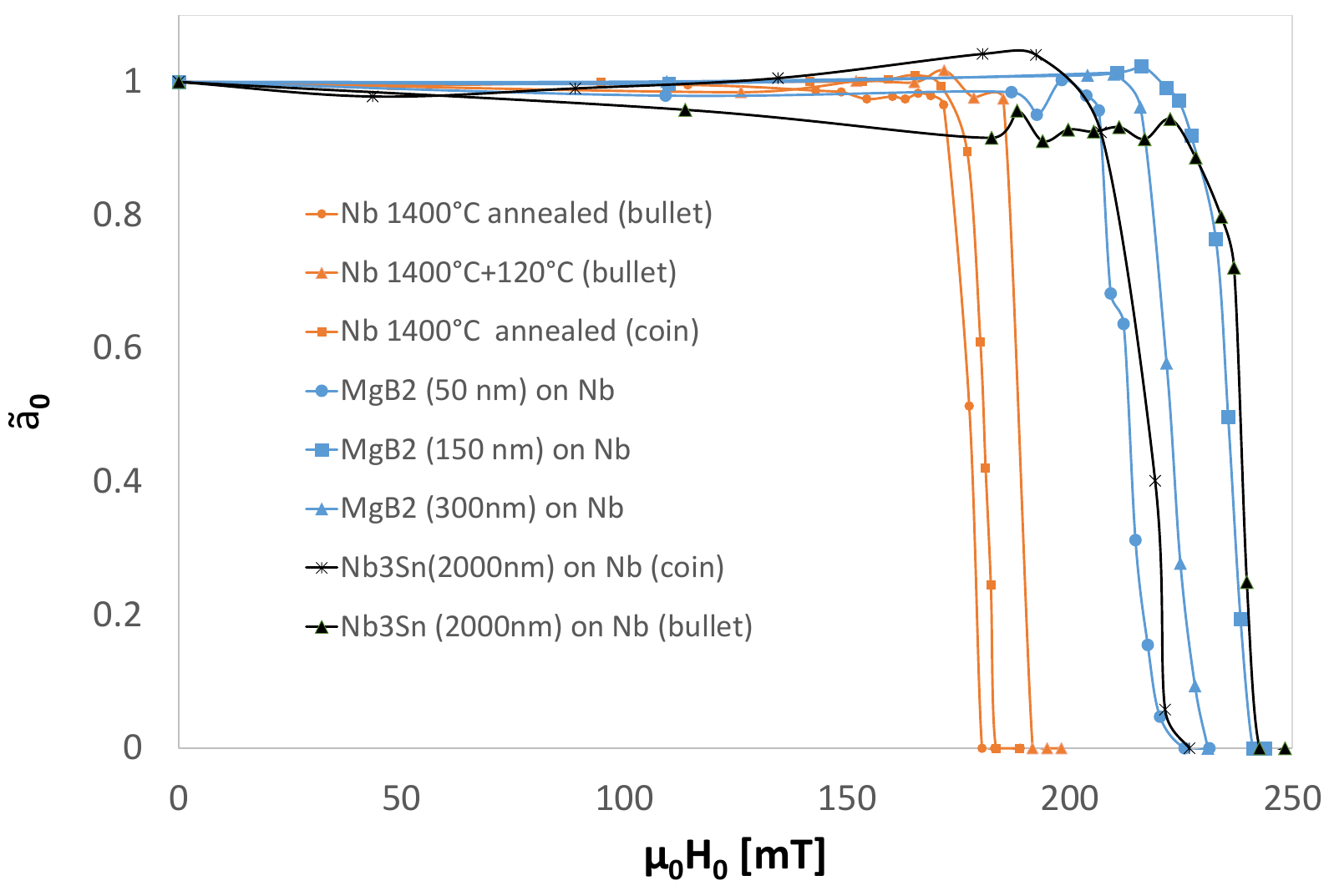}
	\caption{Normalized fit parameter $\widetilde{a}_0$ as a function of $H_0$. }
	\label{fig:Asy}
\end{figure}

Figure \ref{fig:Asy} shows the normalized fit parameter $\widetilde{a}_0$ as a function of the applied field, corrected for geometrical field enhancement, $H\msub{0}$, where $H\msub{0}$/$H\msub{a}$=0.91 and 0.87 for the coin and the ellipsoid respectively \cite{brandt2000superconductors}. For the coin, the geometry is approximated by a long strip with rectangular cross section. Comparing the results of the annealed Nb coin to the bullet shows that this approximation is valid.
Furthermore, the field has been scaled to \unit[0]{K} assuming the empirical relation 
\begin{equation}
H_0(T)=H_0(0\text{K})\left(1-\left(\frac{T}{T\msub{c}}\right)^2\right),
\label{eq:Hc1(T)}
\end{equation}
assuming $T\msub{c}$=\unit[9.25]{K} for Nb \cite{finnemore1966superconducting}. 

The lower estimate for $H\msub{entry}$ is the largest measured $H_0$ for which $\widetilde{a}_0>0.95$ or $a_1$=0 holds. The higher estimate is the smallest $H_0$ for which $\widetilde{a}_0<0.05$ holds. This criteria has been chosen since there have been fluctuations in $\widetilde{a}$ on the order of \unit[10]{\%} for some data sets like the Nb$_3$Sn bullet. This is most likely related to the position and polarization of the incoming muon beam. The uncertainty in $H\msub{entry}$ is the difference in $H_0$ for these two data points plus an additional \unit[1]{\%} to account for additional error sources mainly from potential misalignment.



\begin{figure}
	\centering
		\includegraphics[width=0.90\columnwidth]{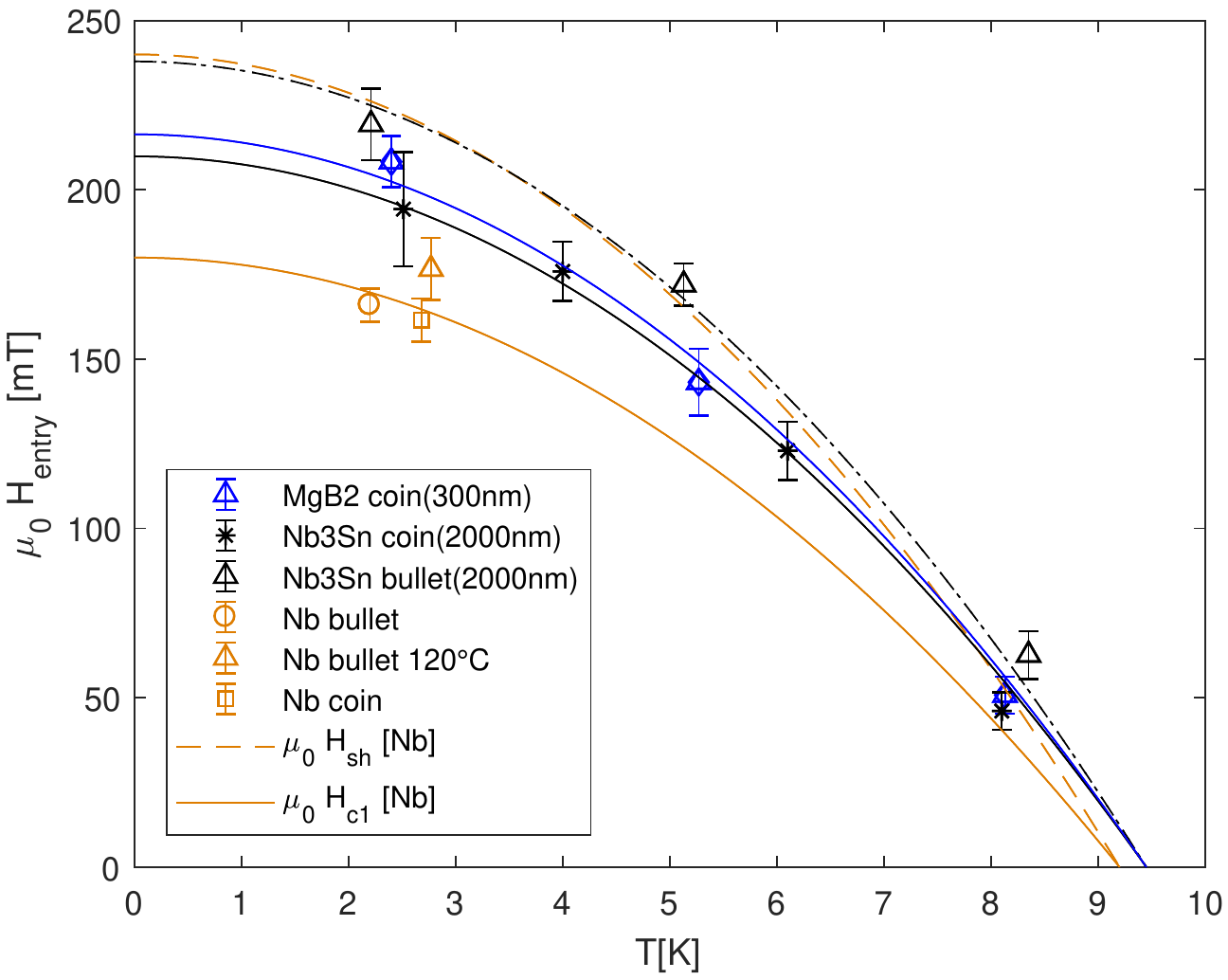}
	\caption{$H\msub{entry}$ as a function of temperature.}
	\label{fig:HentryvsT}
\end{figure}

Figure \ref{fig:HentryvsT} shows $H\msub{entry}$ as a function of temperature for the two Nb$_3$Sn and the \unit[300]{nm} MgB$_2$ sample. The data has been fitted to \refE{eq:Hc1(T)}. Here $T\msub{c}$ was used as a common fit parameter for all three samples. Its value \unit[9.45(0.02)]{K} is slightly above the literature value from Finnemore \unit[9.25]{K} \cite{finnemore1966superconducting} but inconsistent with the much larger critical temperatures of MgB$_2$ and Nb$_3$Sn. $H\msub{entry}(0)$ was fitted individually for each sample. 

For all coated samples, a value significantly above $H\msub{c1$\mid$Nb}$ is found, close to the superheating field of Nb $H\msub{sh}\approx$\unit[240]{mT} \cite{PhysRevB.83.094505}, see \refT{tab:Hentry}. Baking at \unit[120]{$\degree$C} pushes $H\msub{entry}$ from \unit[178(7)]{mT} to \unit[188(4)]{mT}. This can also be correlated to an energy barrier built up at the interface between the dirty layer and the clean bulk as predicted in \cite{checchin2016physics}. Note that \unit[120]{$\degree$C} baking reduces the mean free path at the surface and therefore also $H\msub{c1}$ \cite{Romanenko_muSR}.\\
%
%
\begin{table}
	\centering
			\begin{tabular}{l|c}
			Sample & $\mu_0H\msub{entry}$(0K) \\
			\hline
		Nb \unit[1400]{$\degree$C} (bullet) & 178(7)/- \\
		Nb \unit[1400]{$\degree$C}+\unit[120]{$\degree$C} (bullet) &  188(4)/-\\
		Nb \unit[1400]{$\degree$C} (coin) & 177(7)/- \\
		Nb$_3$Sn bullet & 233(11)/238(43)  \\
		Nb$_3$Sn coin & 210(18)/210(43)\\
		MgB$_2$ 50nm & 216(11)/-   \\
		MgB$_2$ 150nm & 233(9)/- \\
		MgB$_2$ 300nm & 223(9)/216(42)\\
		\end{tabular}
	\caption{$H\msub{entry}$(0K) for all samples used in this study. The first value is from a measurement at about \unit[2.5]{K} and corrected to \unit[0]{K} using \refE{eq:Hc1(T)}. The second one is from a fit to \refE{eq:Hc1(T)} with common $T\msub{c}$. For details see text.}
	\label{tab:Hentry}
\end{table}
%
%
%

%

\section{Simulation}

There is no clear trend in $H\msub{entry}$ vs. layer thickness. This suggests that the superconductor-superconductor (SS) boundary is providing effective shielding up to $H\msub{sh$\mid$ Nb}$, while the superconductor-vacuum (SV) boundary is not providing shielding above its lower critical field $H\msub{c1}$.
%
%
%
%
%
%
%
%
%
Note that realistic surfaces contain defects. A possible explanation why the SS but not the SV boundary provides shielding above $H\msub{c1}$ could be that the proximity effect recovers the order parameter $\psi$ in the vicinity of defects at the SS boundary. 

\begin{figure}
	\centering
		\includegraphics[width=\columnwidth]{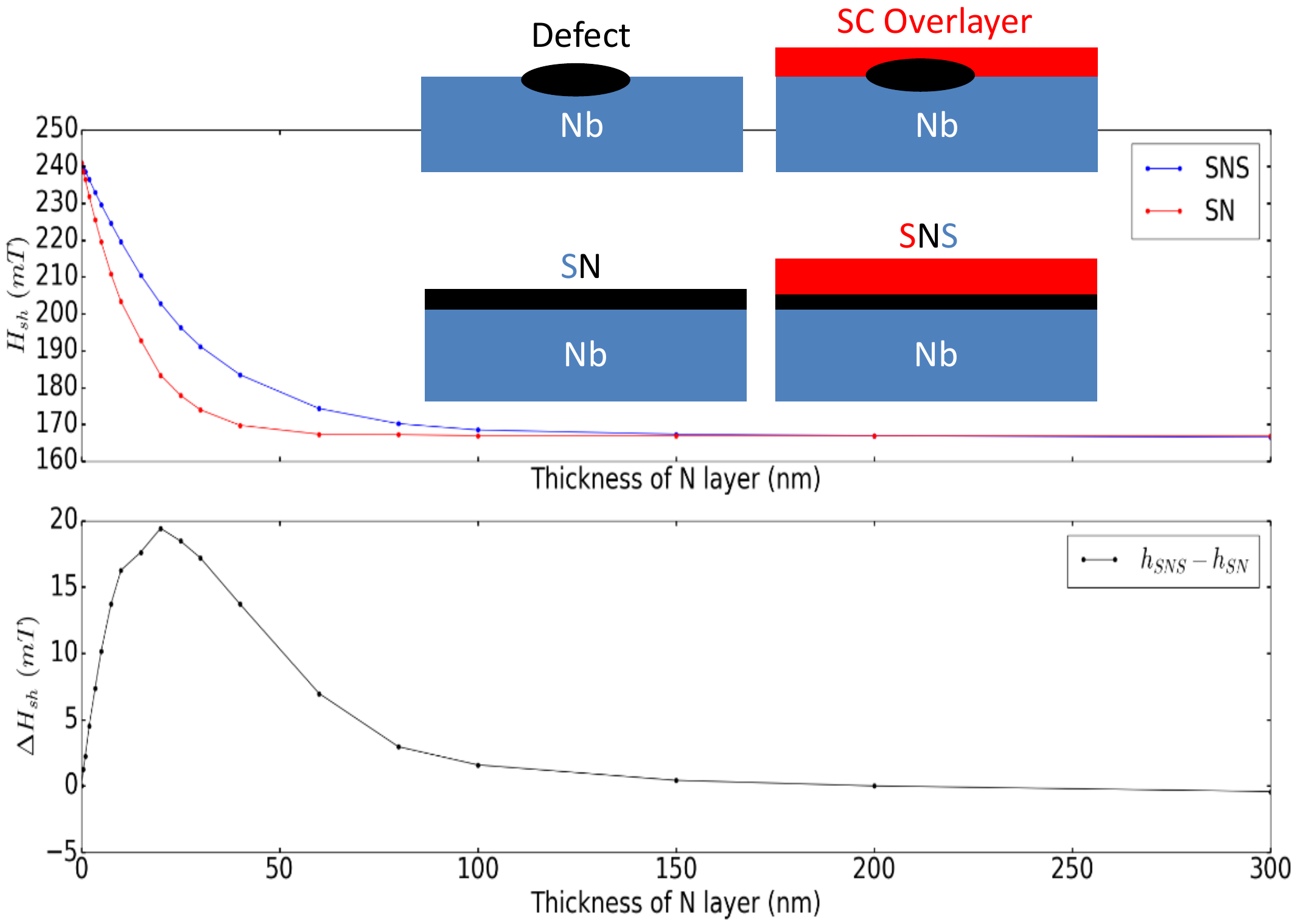}
	\caption{On the top it is shown that for small thicknesses of normal conducting layers, an SNS system should have a higher superheating field than an SN system.  For large thicknesses of normal conducting layers, there is only a depression, as there are essentially just 2 independent SN systems, as Cooper pairs do not travel across the normal conducting layer. The reason for the depressed superheating field for large N is the local depression at the interface from the normal conducting layer, as opposed to a vacuum/insulator layer. On the bottom is $\Delta H=H\msub{SNS}-H\msub{SN}$, as a function of the thickness of the N layer.}  
	\label{fig:SNS}
\end{figure}

In the following we present a simplified simulation model to strengthen support for this hypothesis. It has to be noted that this model cannot serve as a quantitative description of the experimental results, due to several simplifications as outlined in the following.
In the interests of simple calculations, we assume that the defects are approximately the same size, and are very frequent, such that they can be approximated in a 1D model. 
%
The systems of interest are SN and SNS systems comprised of a semi-infinite superconducting slab for $x<0$, a normal conducting plane with finite thickness for $0<x<d$ and for the SNS case, another superconducting slab for $x>d$, see \refF{fig:SNS} (Cartoon on top). Such systems are well studied from the point of view of Josephson junctions. Here the same methods as described in \cite{chapman1995ginzburg} are applied. This treatment uses the non-dimensional Ginzburg-Landau functionals for each domain and minimizes the sum of the functionals with respect to the normalized order parameter $\psi$ and the normalized vector potential $\textbf{A}$. In the superconducting region the dimensionless Ginzburg-Landau equations are (see e.g. \cite{chapman1995ginzburg}):
\begin{equation}
\left( \frac{i}{\kappa} \nabla + \textbf{A} \right)^2 \psi - \psi + \left|\psi \right|^2 \psi = 0,
\end{equation}
\begin{equation}
\left(\nabla\times\nabla\times \textbf{A} \right)=\frac{-i}{2\kappa} \left(\psi^{*} \nabla \psi - \psi \nabla \psi^{*} \right) - \left|\psi \right|^2 \textbf{A},
\end{equation}
where $\kappa=\frac{\lambda}{\xi}$ is the Ginzburg-Landau parameter. To arrive at these equations, the normalizations
\begin{eqnarray}
\widetilde{\psi}=\sqrt{\frac{\mid a \mid}{b}}\psi, &\quad \widetilde{x}=\sqrt{\frac{\widetilde{m} b c^2}{4\pi\mid a \mid e^2 \mu\msub{s}}}x=\frac{x}{\lambda\sqrt{\mu_s}} \\ \nonumber
\widetilde{\textbf{A}}=\sqrt{\frac{2\mid a \mid c^2 \widetilde{m}_s}{e^2}}\textbf{A}, &\quad \widetilde{\textbf{H}}=\sqrt{\frac{8\pi a^2}{b \mu_s}} \textbf{H} \\ \nonumber
\end{eqnarray}
were applied (tilde indicates physical units). $a$ and $b$ are constants, $e\msub{s}$ and $m\msub{s}$ are twice the electron charge and mass respectively, c is the speed of light and $\mu_s$ is the permeability. The modified Ginzburg-Landau equations for the superconducting charge carriers within the normal conducting domain are \cite{chapman1995ginzburg}:
\begin{equation}
\left( \frac{i}{\kappa} \nabla + \textbf{A} \right)^2 \psi + \alpha \psi= 0\, 
\label{eq:GL_N1}
\end{equation}
\begin{equation}
\left(\nabla\times\nabla\times \textbf{A} \right)=\frac{-1}{m_n}\left(\frac{-i}{2\kappa} \left(\psi^{*} \nabla \psi - \psi \nabla \psi^{*} \right) - \left|\psi \right|^2 \textbf{A} \right),
\label{eq:GL_N2}
\end{equation}
where $\alpha=\widetilde{m}\msub{n} a\msub{n}/(\widetilde{m}\msub{s} \mid a\msub{s}\mid) $.
%
%
$\psi$ is taken to be real, which can be done for our problem in one dimension without loss of generalization and simplifies the calculations to a real second order ordinary differential equations boundary value problem, since it results in
\begin{equation}
\psi^{*} \nabla \psi - \psi \nabla \psi^{*}=0. \nonumber
\end{equation}
%
%
%
%
%



The SNS case was examined for identical superconductors only. This was done so as to avoid an issue with \refE{eq:GL_N1} and \refE{eq:GL_N2}, where the parameters used would be dependent on which superconductor the superconducting charge carriers originated from.  Without this assumption, the equations would have to be reformulated to also include the effects of superconducting charge carriers with different normalizations mixing. 
%

The superheating field was taken to be the largest field for which a Meissner state solution could be found. To evaluate this, a magnetic field is applied at the outer surface of the N-layer. This assumes that a vortex has penetrated the SV boundary and reflects the hypothesis that only the SS but not the SV boundary is providing shielding above $H\msub{c1}$. 
With just one type of superconductor, $\psi$ and $\textbf{A}$ are continuous and differentiable at the interfaces. The systems were solved explicitly. For each domain two of the four continuity conditions were forced to be the value on the boundary of the other domain, and the other conditions were allowed to vary. Then, the solutions were iterated until they converged within a tolerance. 
Using thicknesses of 1000 and \unit[1500]{nm} \footnote{These layer thicknesses are used to allow a decay of the wave function to zero, so that the standard boundary conditions $d\psi/dx(x\mapsto\infty)=0$ and curl$\textbf{A}(x\mapsto\infty)=H$ can be used in a finite domain with reasonable calculation time.} for the outer and inner superconductor respectively, $H\msub{sh}$ was obtained for the SN and SNS systems, see \refF{fig:SNS}. The red line shows the depression of $H\msub{sh}$ as a function of normal conducting layer thickness $d$. Mass, charge and permeability were taken to be the same between the normal conducting layer and the superconducting layer. This corresponds to Cooper pairs in a normal conducting plane of Nb on top of superconducting Nb.
%
%
The blue line shows two bulk S layers (the same superconductor with $\kappa$=1.04 for clean Nb \cite{maxfield1965superconducting}), separated by a thin N layer. In both cases, there is no change for thick N layers, which is intuitively expected, as Cooper pairs will only penetrate a distance on the order of the coherence length $\xi$ into the N layer, so beyond that, the bulk SCs are independent. 

For large $d$ and $d\rightarrow 0$, there are no substantial differences found between SN and SNS systems. This can be intuited from an assumption that Cooper pairs penetrate a distance $\approx \xi$ into the normal layer. Then, for a thick layer $(d>>\xi)$, there should be no difference, as the SNS system effectively becomes an SN system.  For a thin layer ($d\rightarrow 0$) there should also be no difference, as both the SN and SNS systems are reduced to a bulk superconductor with no N-layer. For intermediate layers $(d \approx \xi)$, the density of the penetrating Cooper pairs in the N-layer will be greater, resulting in increased shielding.

%
%
%
%
%
%
\section{Conclusion}

In \cite{kubo2016multilayer} it was theoretically shown that there is a second energy barrier at the SS interface of a layered superconductor. However, no reasoning was given under what circumstances this boundary would provide shielding above $H\msub{c1$\mid$substrate }$ and whether this boundary is more stable than the SV boundary. The experimental results and the simulation presented here suggest that the proximity effect can recover the stability of $\psi$ near defects and therefore increase $H\msub{entry}$ up to $H\msub{sh}$ at the SS boundary. However the 1D simulation model is a rather crude approximation of a bi-layer superconductor with defects. A more realistic simulation will require higher dimensions \cite{du2005numerical}. 

In order to be useful to SRF applications a thin overlayer thickness $d<\lambda$ is needed to avoid strong dissipation from vortices within the relevant London layer of a few nm. The results with $d \gg \lambda$ are interesting in the sense that they provide insight in the physics of layered superconductors. Considering the hypothesis that the proximity recovers the stability of $\psi$ near defects, layers with $d \approx \xi\msub{layer}$ could potentially be sufficient to increase $H\msub{entry}$. This is consistent with the increased $H\msub{entry}$ reported here for a \unit[120]{$\degree$C} baked sample and recently reported increased $E\msub{acc}$ for SRF cavities treated with modified low temperature baking recipes \cite{grassellino2017unprecedented}. Furthermore, the proposed hypothesis gives an alternative explanation for recently reported magnetometry measurements on MgB$_2$ on Nb ellipsoidal samples \cite{tan2016magnesium} if pinning is also taken into account. As shown in \cite{Junginger_muSR_Overview} pinning is not negligible for this geometry if the samples are not annealed.  

Further simulations with smaller superconducting layers in higher dimensions and further $H\msub{entry}$ measurements with different layer thicknesses are necessary to further test the proposed hypothesis.   


\section{Acknowledgment}
This research was supported by a Marie Curie International Outgoing Fellowship within the EU Seventh Framework Programme for Research and Technological Development (2007-2013). The authors would like to thank T. Tan, M. Wolak, W.Withanage, X. Xi, D. Hall, S. Posen and M. Liepe for providing the samples used in this study. Thanks to B. Waraich, J. Keir, J. Wong and L. Lambert for producing the bullet samples and performing the heat treatments.

\end{document}